\documentclass[12pt]{iopart}
\usepackage{iopams}
\usepackage{epsf}  

\begin{document}

\title[Mean Field Theory of Thermal Superradiance]
{Thermal Superradiance and the Clausius-Mossotti Lorentz-Lorenz Equations}

\author{S. Sivasubramanian\dag\ 
A. Widom\dag\footnote[3]{To whom correspondence should be 
addressed (a.widom@neu.edu)}\ and Y.N. Srivastava\dag\ddag }
\address{\dag\ Physics Department, Northeastern University, 
Boston MA 02115, USA}
\address{\ddag\ Dipartimento di Fisica \& INFN, 
Universit\'a di Perugia, Perugia, Italia}

\begin{abstract}
Electric polarization phenomena in insulating systems 
have long been described in mean field theory 
by the (static) Clausius-Mossotti or (dynamic) Lorentz-Lorenz 
polarizabilities. It is here shown, in the strong coupling  
regime, that a thermodynamic phase instability exists in these models. 
The resulting thermodynamic phase diagram coincides with that obtained from 
Dicke-Preparata model of thermal superradiance.   
\end{abstract}
\pacs{42.50.Fx, 32.10.Dk, 33.15.Kr, 42.65.An}
\vspace{.5cm}

A property of fundamental importance in the physics and chemistry 
of condensed matter insulators is the dielectric response function  
\begin{equation}
\varepsilon =1+4\pi \chi 
\end{equation}
In principle, \begin{math} \varepsilon  \end{math} is largely 
determined by the polarizability[1-12]  \begin{math} \alpha \end{math} 
of the constituent molecules. Precisely relating  
\begin{math} \varepsilon \end{math} to \begin{math} \alpha \end{math}   
is a statistical thermodynamic problem of considerable complexity. 
In the lowest order mean field theory approximation, the problem is 
solved by the Clausius-Mossotti equation[13-16] 
\begin{equation}
\left(\frac{4\pi \alpha }{3v}\right)
=\left(\frac{\varepsilon -1}{\varepsilon +2}\right),
\end{equation}
where \begin{math} v \end{math} is the volume per molecule. 
Equivalently, the mean field Clausius-Mossotti 
electric susceptibility is then 
\begin{equation}
\chi=\left(\frac{(\alpha /v)}{1-(4\pi \alpha /3v)}\right).
\end{equation}
Note that the stability of Clausius-Mossotti mean field theory 
requires the inequality for the static polarizability  
\begin{equation}
4\pi \alpha < 3v\ \ \ {\rm (stable\ normal\ phase)}.
\end{equation}
When the above inequality is violated, the normal phase 
becomes unstable and a new stable phase appears. Our 
purpose is to show that this new stable phase is nothing but the 
superradiant phase implicit in superradiant[17-26] Dicke-Preparata 
models. The derivation of the superradiant phase 
diagram is considerably simplified. To see what is involved, 
one may consider the response functions 
\begin{math} \alpha \end{math}, \begin{math} \chi \end{math} 
and \begin{math} \varepsilon \end{math} generalized to 
finite frequency \begin{math} \omega \end{math}. 

When the response functions  are contemplated  
at finite frequency, for example in the index of refraction 
\begin{math} n(\omega )=\sqrt{\varepsilon (\omega )}\ \end{math}, 
the Clausius-Mossotti Eqs.(1), (2) and (3) become the Lorentz-Lorenz 
equations[27-30]. 
These have a surprising feature regarding frequency shifts. 
For example, suppose that a single atom with atomic number 
\begin{math} Z \end{math} has a strong resonance at frequency 
\begin{math} \omega_\infty \end{math}, i.e. with 
\begin{math} m  \end{math} as the electron mass,   
\begin{equation}
\alpha (\omega )\approx \left(\frac{Ze^2}{m}\right)
\frac{1}{\omega_\infty ^2-\omega ^2-2i\gamma \omega },
\end{equation}
Eqs.(3) and (5) then imply 
\begin{equation}
\chi(\omega )\approx \left(\frac{Ze^2}{vm}\right)
\frac{1}{\omega_0 ^2-\omega ^2-2i\gamma \omega },
\end{equation}
with a ``renormalized'' resonant frequency 
\begin{math} \omega_0 \end{math} given by 
\begin{equation}
\omega_0^2=\omega_\infty ^2- \left(\frac{4\pi Ze^2}{3vm}\right)
=\omega_\infty ^2\left\{1-
\left(\frac{4\pi \alpha }{3v}\right)\right\}, 
\end{equation}
where \begin{math} \alpha \equiv \alpha (\omega =0)  \end{math} is 
the static polarizability. The {\em macroscopic} Lamb frequency shift 
\begin{math} \Delta \omega = \omega_\infty -\omega_0 \end{math} 
is certainly worthy of note. It is due to the coherent collective 
oscillation of a large number of molecular dipole moments.

Electric dipolar motions in (say) a single atom move at a Bohr frequency 
\begin{math} \omega_\infty \end{math}
determined by the energy difference between two states 
\begin{math} \left|i\right> \end{math} and 
\begin{math} \left|f\right> \end{math} connected by a dipole 
matrix element 
\begin{math} \left<f\right|{\bf p} \left|i\right> \ne 0\end{math}, i.e. 
\begin{math} \hbar \omega_\infty =(\epsilon_f-\epsilon_i) \end{math}. 
In a condensed matter sample of similar atoms, 
the dipolar motions in one atom radiate an electromagnetic field
which can drive a neighboring atomic dipole into oscillation. The 
motion can become collective. In the {\em mean field} 
theory of Clausius-Mossotti and Lorentz-Lorenz, the 
driving collective macroscopic electric field 
\begin{math} \bar{\bf E} \end{math} is related to the dipole 
moment per unit volume \begin{math} {\bf P} \end{math} via 
\begin{math} \bar{\bf E}=-(4\pi /3){\bf P} \end{math}. The 
mean electric field phase locks the oscillations of many atoms. 
The collective motions of the large number of dipoles yields 
the macroscopic Lamb shift frequency renormalization 
\begin{math} \omega_\infty \to \omega_0 \end{math} 
in accordance with Eq.(7). Mean field 
\begin{math} \bar{\bf E} \end{math} theory thereby implies  
a new collective Bohr frequency which from a quantum mechanical 
viewpoint arises from more closely spaced macroscopic energy levels
\begin{math} \hbar \omega_0 =(E_f-E_i)<\hbar \omega_\infty 
=(\epsilon_f-\epsilon_i) \end{math}. 

At the critical temperature \begin{math} T_c  \end{math} for stability, 
we have from the static polarizability stability Eq.(4) the condition 
\begin{equation}
\left(\frac{4\pi \alpha (T_c)}{3}\right)=v 
\ \ \ ({\rm critical\ point}),
\end{equation} 
where \begin{math} \alpha_c=\alpha (T_c) \end{math}. 
From the dynamical Eq.(7) viewpoint, the renormalized frequency 
is lowered to zero at the critical temperature  
\begin{math} \lim_{T\to T_c +}\{\omega_0(T)\}=0 \end{math} 
which again leads to Eq.(8). 

In order to compute \begin{math} \alpha (T) \end{math} one 
needs a model Hamiltonian \begin{math} H_{mol} \end{math} for 
a single molecule. In the presence of an external electric field 
\begin{math} {\bf F} \end{math} one may then compute the single 
molecule partition function 
\begin{equation}
\Phi ({\bf F},T)=-k_BT\ln\left(Tr\ 
e^{-(H_{mol}-{\bf p\cdot F})/k_BT}\right).
\end{equation}
where \begin{math} {\bf p} \end{math} is the electric dipole 
operator of the molecule. The mean dipole moment is determined 
by 
\begin{equation}
d\Phi =-sdT-\bar{\bf p}\cdot d{\bf F},
\end{equation}
and the polarizability of the molecule is given by
\begin{equation}
\alpha_{ij}=\lim_{|{\bf F}|\to 0}
\left(\frac{\partial \bar{p}_i}{\partial F_j}\right)_T .
\end{equation}

For a ``two level atom'' corresponding to the Dicke model, 
one chooses the model Hamiltonian 
\begin{equation}
H_2-{\bf p}_2{\bf \cdot F}=
\pmatrix{-\Delta & -\mu F \cr
-\mu F & \Delta }
\end{equation}
yielding
\begin{equation}
\Phi_2 =-k_BT \ln 2-k_BT\ln
\left\{
\cosh\left(\frac{\sqrt{\Delta^2+\mu^2F^2}}{k_BT}\right)
\right\}
\end{equation}
so that 
\begin{equation}
\alpha_2 (T) =\left(\frac{\mu^2}{\Delta }\right)
\tanh\left(\frac{\Delta }{k_BT}\right).
\end{equation}
Employing Eqs.(8) and (14) and the definition of the critical volume 
per two-state atom 
\begin{equation}
v_c=\left(\frac{4\pi \mu^2}{3\Delta }\right)
\end{equation} 
we find for the critical temperature of the  Clausius-Mossotti model 
for two state atoms 
\begin{equation}
k_BT_{c2}=\left(\frac{2\Delta }
{\ln\{(v_c+v)/(v_c-v)\}}\right).
\end{equation}
The above Eq.(16) is identical to the critical temperature of the Dicke 
model as derived in previous work[31, 32].

For atoms in an electronic zero angular momentum  
(\begin{math} J=0 \end{math}) state, it is more realistic to consider 
a four state model model with three (\begin{math} J=1 \end{math}) 
degenerate states (\begin{math} M_J=0,\pm 1 \end{math}). 
The resulting Hamiltonian four state matrix may then be written as 
\begin{equation}
H-{\bf p \cdot F}=
\pmatrix{-\Delta & -\mu F_x & -\mu F_y & -\mu F_z \cr 
-\mu F_x & \Delta & 0 & 0 \cr
-\mu F_y & 0 & \Delta & 0 \cr 
-\mu F_z & 0 & 0 & \Delta }.
\end{equation} 
The free energy implicit in the Hamiltonian matrix of Eq.(17) 
may be written as 
\begin{equation}
\Phi=-k_BT\ln 2-k_BT\ln 
\left\{
e^{-\Delta /k_BT}+
\cosh\left(\frac{\sqrt{\Delta^2+\mu^2|{\bf F}|^2}}{k_BT}\right)
\right\}
\end{equation}
so that 
\begin{equation}
\alpha (T) =\left(\frac{\mu^2}{\Delta }\right)
\frac{\sinh(\Delta /k_BT)}{\exp(-\Delta /k_BT)+\cosh(\Delta /k_BT)}\ .
\end{equation}
Employing Eqs.(8), (15) and (19) yields the critical temperature 
\begin{math} T_c \end{math} for the superradiant phase transition  
found by solving
\begin{equation}
\frac{\sinh(\Delta /k_BT_c)}{\exp(-\Delta /k_BT_c)+\cosh(\Delta /k_BT_c)}
=\left(\frac{v}{v_c}\right).
\end{equation} 
The phase diagrams implicit in both Eqs.(16) and (20) are shown in Fig.1.
The difference between the two level model and the four level model is 
merely a shift in the phase boundary. The existence of the superradiant 
phase is robust with respect to changes in the detailed model 
structure.

\begin{figure}[!t]
\noindent
\epsfxsize=296pt
\centerline{\epsffile{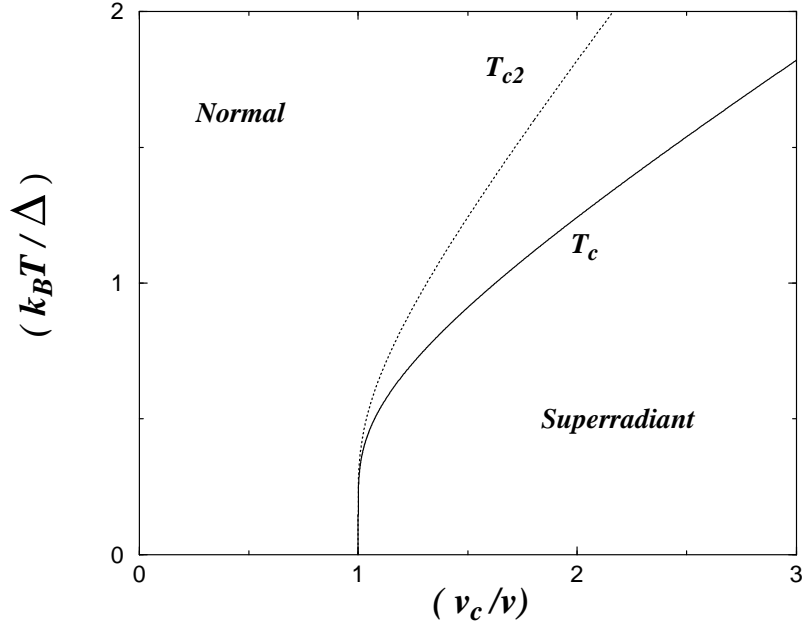}}
\vspace{4pt}
\caption{\label{PhaseDiagram}Shown as a ``solid curve'' is the critical 
temperature $T_c$ for the four level atom as computed from Eq.(20). 
The curve divides the $(T,v)$ plane into regions which are either 
superradiant or normal. The ``dotted curve'' shows the critical 
temperature which follows from the 
superradiant\cite{Siva_2001_1,Siva_2001_2} Dicke 
two level atom model.}
\end{figure}                   

We have developed a very simple algorithm for finding 
the regions in the  \begin{math} (T,v) \end{math} plane 
for which the superradiant phase exists. Linear magnetic 
materials can be either paramagnetic or diamagnetic. 
Linear electrical systems can only occur in nature as 
paraelectrics. Diaelectric systems can be ruled out\cite{LandL}. 
It is the absence of diaelectricity which accounts for the 
reliability of the algorithm. The computation of the phase diagram 
requires (i) a reliable computation of the single molecule 
polarizability, and (ii) the dielectric constant of the bulk 
material. The critical temperature has the form of an implicit 
equation    
\begin{equation}
\alpha (T_c)=g(v,T_c)\ \ {\rm where} 
\ \ g_{CM}=\left(\frac{3v}{4\pi }\right).
\end{equation}
The Clausius-Mossotti function \begin{math} g_{CM} \end{math} 
is valid in mean field theory. The fluctuation corrections to 
Clausius-Mossotti mean field theory[34-39], (i.e. the precise  
\begin{math} g(v,T) \end{math} function) may be computed 
systematically via perturbation theory. The corrections are merely 
quantitative. Qualitatively, the superradiant phase is shifted but 
left intact.

\end{document}